\documentclass[seceq]{ptptex}

\usepackage{graphicx}

\pubinfo{\tt Vol.~114 (3), 533-538 (Sept. 2005), math-ph/0504055 v3}

\title{
NONLINEAR SECOND ORDER ODE'S:\\ FACTORIZATIONS AND PARTICULAR SOLUTIONS
}

\author{       
O. \textsc{Cornejo-P\'erez and H. C. \textsc{Rosu}}%
}

\inst{     
Potosinian Institute of Science and Technology\\
Apdo Postal 3-74 Tangamanga, 78231 San Luis Potos\'{\i}, Mexico
}

\abst{         
We present particular solutions for the following important nonlinear second order
differential equations: modified Emden, generalized Lienard, convective Fisher, and generalized Burgers-Huxley. For the latter two equations these solutions are obtained in the travelling frame.
All these particular solutions are the result of extending a simple and efficient factorization method
that we developed in Phys. Rev. E {\bf 71} (2005) 046607.
}

\begin{document}

\maketitle

\section{Introduction}

\noindent
The purpose of this paper is to obtain, through the factorization technique, particular solutions of the following type of differential equations
\begin{equation}
{\ddot u} + g(u){\dot u}+ F(u)=0~,
\label{ec1}
\end{equation}
where the dot means the derivative $D=\frac{d}{d\tau}$, and $g(u)$ and $F(u)$ could in principle be arbitrary functions of $u$.
This is a generalization of what we did in a recent paper for the simpler equations
with $g(u)=\gamma$, where $\gamma$ is a constant parameter \cite{rosu1}.
Factorizing Eq.~(\ref{ec1})  means to write it in the form
\begin{equation}
\left[ D-\phi_{2}(u)\right]\left[ D-\phi _{1}(u)\right]u=0~.
\label{ec2}
\end{equation}
Performing the product of differential operators leads to the equation
\begin{equation}
{\ddot u}-\frac{d\phi _{1}}{du}u{\dot u}-\phi
_{1}{\dot u} -\phi_{2}{\dot u}+\phi _{1}\phi_{2}u=0\, ,
\label{ec3}
\end{equation}
for which one very effective way of grouping the terms is\cite{rosu1}
\begin{equation}
{\ddot u}-\left(\phi _{1}+\phi_{2}+\frac{d\phi
_{1}}{du}u\right){\dot u} +\phi _{1}\phi_{2}u=0\, .\label{ec4}
\end{equation}
Identifying Eqs. (\ref{ec1}) and (\ref{ec4}) leads to the conditions
\begin{eqnarray}
g(u)&=&-\left(\phi _{1}+\phi_{2}+\frac{d\phi _{1}}{du}u\right)\label{ec5}\\
F(u)&=&\phi _{1}\phi_{2}u~.\label{ec6}
\end{eqnarray}
If $F(u)$ is a polynomial function, then $g(u)$ will have the same
order as the bigger of the factorizing functions
$\phi _{1}(u)$ and $\phi _{2}(u)$, and will also be a function of the
constant parameters that enter in the expression of $F(u)$.



 In this research, we extend the method to the following cases:
the modified Emden equation, the generalized Lienard
equation, the convective Fisher equation, and the generalized
Burgers-Huxley equation. All of them have significant applications
in nonlinear physics and it is quite useful to know their explicit particular solutions.
The present work is a detailed contribution to this issue.

\section{Modified Emden equation}

\noindent
We start with the modified Emden equation with cubic nonlinearity that has been most recently discussed by Chandrasekhar {\em et al}\cite{vchandra},
\begin{equation}
{\ddot u} + \alpha u{\dot u} + \beta u^{3}=0~. \label{ec7}
\end{equation}
{\bf 1}) 
$\phi _{1}(u)=a_{1}\sqrt{\beta}u~,\quad 
\phi_{2}(u)=a_{1}^{-1}\sqrt{\beta}u$, 
($a_{1}\neq 0$ is an arbitrary constant).

\noindent
Then Eq. (\ref{ec5})
leads to the following form of the function $g(u)$
\begin{equation}
g_1(u)=
-\sqrt{\beta}\left(\frac{2a_{1}^{2}+1}{a_{1}} \right)u~.
\label{ec9}
\end{equation}
Thus we can identify
$\alpha=-\sqrt{\beta}\left(\frac{2a_{1}^{2}+1}{a_{1}} \right)$, or
$a_{1_{\pm}}=\frac{-\alpha\pm\sqrt{\alpha^{2}-8\beta}}{4\sqrt{\beta}}$,
where we use $a_{1}$ as a fitting parameter providing that
$a_{1}<0$ for $\alpha>0$. 
Eq. (\ref{ec7}) is now rewritten as 
\begin{equation}
{\ddot u} - \sqrt{\beta}\left(2a_{1}+a_{1}^{-1}\right)
u{\dot u} + \beta u^{3}\equiv\left( D- a_{1}^{-1}\sqrt{\beta}u\right)\left( D-
a_{1}\sqrt{\beta}u\right)u=0~. \label{ec10}
\end{equation}
Therefore, the compatible first order differential equation is
${\dot u}-a_{1}\sqrt{\beta}u^{2}=0$, whose
integration gives the particular solution of
Eq. (\ref{ec10})
\begin{equation}
u_1=-\frac{1}{a_{1}\sqrt{\beta}(\tau-\tau_{0})} \quad {\rm or} \quad u_1=\frac{4}{(\alpha \pm \sqrt{\alpha^2
-8\beta})(\tau-\tau_{0})}~,\label{ec13}
\end{equation}
where $\tau_{0}$ is an integration constant.

\bigskip
\noindent
{\bf 2}) 
$\phi _{1}(u)=a_{1}\sqrt{\beta}u^2,\quad
\phi_{2}(u)=a_{1}^{-1}\sqrt{\beta}$.
Then, one gets
\begin{equation}
g_2(u)=-\sqrt{\beta}\left(a_{1}^{-1}+3a_{1}u^{2} \right)~.
\end{equation}
Therefore, $g_2$ is quadratic being higher in order than the linear $g$ of the modified Emden equation. We thus get
the particular case $GE=3\beta$, $A=0$ of the Duffing-van der Pol equation (see case {\bf 3} of the next section)
\begin{equation}
{\ddot u} - \sqrt{\beta}\left(a_{1}^{-1}+3a_{1}u^{2}
\right){\dot u} + \beta u^{3}\equiv \left(D-a_{1}^{-1}\sqrt{\beta} \right)\left( D-
a_{1}\sqrt{\beta}u^{2}\right)u =0~,\label{ec15}
\end{equation}
which leads to the compatible first order differential equation
${\dot u}-a_{1}\sqrt{\beta}u^{3}=0$
with the solution
\begin{equation}
u_2=\frac{1}{[-2a_{1}\sqrt{\beta}(\tau-\tau_{0})]^{1/2}}~.\label{ec18}
\end{equation}


\section{Generalized Lienard equation}
Let us consider now the following generalized Lienard equation
\begin{equation}
{\ddot u} + g(u){\dot u} + F_3=0~,\label{ec19}
\end{equation}
where $F_3(u)=Au+Bu^{2}+Cu^{3}$. We introduce the notation
$\Delta =\sqrt{B^{2}-4AC}$,
and assume that $\Delta ^2>0$ holds. Then:


\medskip
\noindent
{\bf 1}) 
\noindent
$\phi _{1}(u)=a_{1}\left(\frac{\left(B+\Delta\right)}{2}+Cu\right)~,\quad
\phi_{2}(u)=a_{1}^{-1}\left(\frac{\left(B-\Delta\right)}{2C}+u\right)$;
$g(u)$ takes the form
\begin{equation}
g_1(u)=-\left[\frac{\left(B+\Delta\right)}{2}\,a_1+\frac{\left(B-\Delta\right)}{2C}\,a_{1}^{-1}+\left(
2Ca_{1}+a_{1}^{-1} \right)u\right]~.
\end{equation}
For $g(u)=g_1(u)$, we can factorize Eq.~ (\ref{ec19}) in the form
\begin{equation}
\left[ D- a_{1}^{-1}\left(\frac{\left(B-\Delta\right)}{2C}+u\right)\right]\left[ D-
a_{1}\left(\frac{\left(B+\Delta\right)}{2}+Cu\right)\right]u =0~.
\label{ec23}
\end{equation}
Thus, from the compatible first order differential equation
${\dot u}-a_{1}(\frac{\left(B+\Delta\right)}{2}+Cu)u=0$,  
the following solution is obtained
\begin{equation}
u_1=
\frac{\left(B+\Delta\right)}{2}
\left({{\textrm{exp}\bigg[-a_{1}\left(\frac{\left(B+\Delta\right)}{2}\right)(\tau-\tau_{0})\bigg]-C}}\right)^{-1}
~.
\label{ec25}
\end{equation}

\medskip
\noindent
{\bf 2})
$\phi _{1}(u)=a_{1}(A+Bu+Cu^2)~,\quad
\phi_{2}(u)=a_{1}^{-1}$; $g(u)$ is of the form
\begin{equation}
g_2(u)=-\left[(a_1A+a_{1}^{-1})+2a_1Bu+3a_1Cu^2\right]~.
\end{equation}
Thus, the factorized form of the Lienard equation will be
\begin{equation}
\left[D-a_{1}^{-1}\right]\left[D-a_1\frac{F_3(u)}{u}\right]u=0
\end{equation}
and therefore we have to solve the equation $\dot{u}-a_1F_3(u)=0$,
whose solution can be found graphically from
\begin{equation}
a_1(\tau - \tau _0)=\ln \left(\frac{u^3}{F_3(u)}\right)^{\frac{1}{2A}}-\ln \left(\frac{2Cu+B-\Delta}{2Cu+B+\Delta}\right)^{\frac{1}{2A}\frac{B}{\Delta}}~.
\end{equation}

\noindent
{\bf 3}) {\em The case $B=0$ and $C=1$: Duffing-van der Pol equation}

\noindent
The $B=0$, $C=1$ reduction of terms in Eq.
(\ref{ec19}) allows an analytic calculation of particular
solutions for the so-called autonomous Duffing-van der Pol
oscillator equation \cite{chandrasekar}
\begin{equation}
{\ddot u} + (G+Eu^2){\dot u} + Au + u^{3}=0~,\label{ec26}
\end{equation}
where $G$ and $E$ are arbitrary constant parameters. Since we want to compare our solutions with those of Chandrasekar {\em et al} \cite{chandrasekar},
we use the second Lienard pair of factorizing functions 
$\phi _{1}(u)=a_{1}(A+u^{2})$ and
$\phi_{2}(u)=a_{1}^{-1}$.
Then
\begin{equation}
g_2(u)=-\left(Aa_1+a_{1}^{-1} + 3a_{1}u^{2}
\right)~.
\label{gpart}
\end{equation}
Eq. (\ref{ec26}) is now rewritten
\begin{equation}
{\ddot u} - \left(a_{1}A+a_{1}^{-1} + 3a_{1}u^{2}
\right){\dot u} + Au + u^{3}\equiv \left[D -a_{1}^{-1} \right] \left[
D-a_{1}(A+u^{2})\right]u=0~.\label{ec28}
\end{equation}
Therefore, the compatible first order equation
${\dot u}-a_{1}(A+u^{2})u=0$ leads by
integration to the particular solution of
Eq. (\ref{ec28})
\begin{equation}
u=\pm\left(
\frac{A\textrm{exp}[2a_{1}A(\tau-\tau_{0})]}{1-\textrm{exp}[2a_{1}A(\tau-\tau_{0})]}
\right)^{1/2}=\pm\left(
\frac{A\textrm{exp}[-\frac{2}{3}AE(\tau-\tau_{0})]}{1-\textrm{exp}[-\frac{2}{3}AE(\tau-\tau_{0})]}
\right)^{1/2}  ~,\label{ec31}
\end{equation}
where the last expression is obtained from the comparison of Eqs. (\ref{ec26}) and (\ref{ec28}) that gives $a_{1}=-\frac{E}{3}$
and $G=\frac{AE^2+9}{3E}$.

This is a more general result for the particular solution than
that obtained through other means by Chandrasekar \textit{et al}
\cite{chandrasekar} that corresponds to $E=\beta$ and $A=\frac{3}{\beta^2}$.

\section{Convective Fisher equation}

Sch\"onborn {\em et al} \cite{schonborn} discussed the following convective Fisher equation
\begin{equation}
\frac{\partial u}{\partial
t}=\frac{1}{2}\frac{\partial^{2}u}{\partial x^{2}}+u(1-u)-\mu
u\frac{\partial u}{\partial x}~, \quad {\rm or} \quad \ddot{u} + 2(\nu -\mu u)\dot{u} +
2u(1-u)=0~,\label{ec32}
\end{equation}
where the transformation to the travelling variable
$\tau=x-\nu t$ was performed in the latter form.
The positive parameter $\mu$ serves to tune the
relative strength of convection.

\medskip
\noindent
{\bf 1}) $\phi _{1}(u)=\sqrt{2}a_{1}(1-u),\quad\phi_{2}(u)
=\sqrt{2}a_{1}^{-1}$. Then
$g(u)=-\sqrt{2}\left([a_{1}+a_{1}^{-1}] -2a_{1}u\right)$.
Therefore, for this $g(u)$,  we can rewrite the ordinary differential form in Eq.~(\ref{ec32}) as
\begin{equation}
\ddot{u} + 2\left(-\frac{1}{\sqrt{2}}(a_{1}+a_{1}^{-1})
+\sqrt{2}a_{1}u \right)\dot{u} + 2u(1-u)=0~.\label{ec35}
\end{equation}
If we set the fitting parameter $a_{1}=-\frac{\mu}{\sqrt{2}}$,
then we obtain $\nu=\frac{\mu}{2}+{\mu}^{-1}$. Eq. (\ref{ec35}) is
factorized in the following form
\begin{equation}
\left[D -\sqrt{2}a_{1}^{-1} \right] \left[
D-\sqrt{2}a_{1}(1-u)\right]u=0~,\label{ec36}
\end{equation}
that provides the compatible first order equation
$
\dot{u}+\mu u(1-u)=0$, whose integration gives
\begin{equation}
u_1=
\left(1\pm \textrm{exp}[\mu (\tau-\tau_{0})]\right)^{-1}~.\label{ec38}
\end{equation}\\
\medskip
\noindent {\bf 2}) Since we are in the case of a quadratic
polynomial, a second factorization means exchanging $\phi _{1}(u)$
and $\phi _{2}(u)$ between themselves. This leads to a convective
Fisher equation with compatibility equation
$\dot{u}-\sqrt{2}a_{1}^{-1}u=0$, where now $a_1=-\sqrt{2}\mu$,
having exponential solutions of the type
\begin{equation}
u_2=
\pm \textrm{exp}[-\mu ^{-1} (\tau-\tau_{0})]~.\label{ec38b}
\end{equation}

\section{Generalized Burgers-Huxley equation}

In this section we obtain particular solutions for the generalized
Burgers-Huxley equation discussed by Wang \textit{et al}
\cite{wang}
\begin{equation}
\frac{\partial u}{\partial t}+\alpha u^{\delta}\frac{\partial
u}{\partial x} - \frac{\partial^{2}u}{\partial x^{2}}=\beta
u(1-u^{\delta})(u^{\delta}-\gamma)~,\label{ec39}
\end{equation}
or in the variable $\tau=x-\nu t$
\begin{equation}
\ddot{u} + (\nu - \alpha u^{\delta})\dot{u}+\beta
u(1-u^{\delta})(u^{\delta}-\gamma)=0~.
\label{ec40}
\end{equation}

\medskip
\noindent
{\bf 1})
$\phi _{1}(u)=\sqrt{\beta}a_{1}(1-u^{\delta})~,\quad\phi_{2}(u)
=\sqrt{\beta} a_{1}^{-1}(u^{\delta}-\gamma)~.$
Then, one gets
\begin{equation}
g_1(u)=\sqrt{\beta}\left(\gamma a_{1}^{-1}-a_{1}
+[a_{1}(1+\delta)-a_{1}^{-1}]u^{\delta}\right)
\label{gbh1}
\end{equation}
and the following identifications of the constant parameters
$\nu=-\sqrt{\beta}\left(a_1-\gamma a_{1}^{-1}\right)$, \quad 
$\alpha=-\sqrt{\beta}\left(a_{1}(1+\delta)-a_{1}^{-1}\right)$. 
Writing Eq. (\ref{ec40}) in factorized form
\begin{equation}
\left[D - \sqrt{\beta}a_{1}^{-1}(u^{\delta}-\gamma)\right]
\left[ D - \sqrt{\beta}a_{1}(1-u^{\delta})
\right]u=0~,\label{ec43}
\end{equation}
the solution
\begin{equation}
u_1=\left( 
1\pm
\textrm{exp}[-a_{1}\sqrt{\beta}\delta(\tau-\tau_{0})] \right)^{-1/
\delta} \label{ec45}
\end{equation}
of the compatible first order equation
$\dot{u}-\sqrt{\beta}a_{1}u(1-u^{\delta})=0$ 
is also a particular kink solution of Eq. (\ref{ec40}). It is easy to solve
the second identification equation for
$a_{1}=a_{1}(\alpha,\beta,\delta)$ leading to
\begin{equation}
a_{1_{\pm}}=\frac{-\alpha\pm\sqrt{\alpha^2+4\beta(1+\delta)}}{2\sqrt{\beta}(1+\delta)}~.\label{ec45a}
\end{equation}
Then Eq. (\ref{ec45}) becomes a function
$u=u(\tau;\alpha,\beta,\delta)$, and
$\nu=\nu(\alpha,\beta,\gamma,\delta)$.

\bigskip
\noindent
{\bf 2})
$\phi
_{1}(u)=\sqrt{\beta}e_{1}(u^{\delta}-\gamma),\quad \phi_{2}(u)
=\sqrt{\beta} e_{1}^{-1}(1-u^{\delta})$.
This pair of factorizing functions lead to
\begin{equation}
g_2(u)=\sqrt{\beta}\left(\gamma e_1
-e_{1}^{-1} +[e_{1}^{-1}-e_{1}(1+\delta)]u^{\delta}\right)
\label{g2bh}
\end{equation}
and the $\nu$ and $\alpha$ identifications:
$\nu=\sqrt{\beta}\left(e_{1}\gamma -e_{1}^{-1}\right)$,  
$\alpha=\sqrt{\beta}\left(e_1^{-1}-e_{1}(1+\delta)\right)$.

\noindent
Eq. (\ref{ec40}) is then factorized in the different form
\begin{equation}
\left[D - \sqrt{\beta}e_{1}^{-1}(1-u^{\delta}) \right] \left[
D -\sqrt{\beta}e_{1}(u^{\delta}-\gamma)
 \right]u=0~.
\label{ec47}
\end{equation}
The corresponding compatible first order equation is now
$\dot{u}-\sqrt{\beta}e_{1}u(u^{\delta}-\gamma)=0$,
and its integration gives a different particular solution of Eq.
(\ref{ec40}) with respect to that obtained for the first choice of
factorizing brackets: 
\begin{equation}
u_2= \left( \frac{\gamma}{1\pm \textrm{exp}[
e_{1}\sqrt{\beta}\gamma\delta(\tau-\tau_{0})]} \right)^{1/
\delta}~. \label{ec49}
\end{equation}
$u_2$ is different of $u_1$ because the
parameter $\alpha$ has changed for the second factorization. Solving the $\alpha$ identification for
$e_{1}=e_{1}(\alpha,\beta,\delta)$ allows to express
the solution given by Eq. (\ref{ec49}) in terms of the parameters of the equation,
$u=u(\tau;\alpha,\beta,\gamma,\delta)$, and also one gets
$\nu=\nu(\alpha,\beta,\gamma,\delta)$.
If we set $\delta =1$ in Eq. (\ref{ec49}),
then from $\alpha=\sqrt{\beta}(e_1^{-1}-2e_{1})$ one can get
$e_{1_{\pm}}=\frac{\alpha\pm\sqrt{\alpha^2+8\beta}}{4\sqrt{\beta}}$ that can be used to obtain
$\nu _{\pm}=\nu(\alpha,\beta,\gamma)$.
The solutions given by Eqs.~(\ref{ec45}) and (\ref{ec45a}) and in  (\ref{ec49})
have been obtained previously
by Wang \textit{et al} \cite{wang} by a different procedure.

\section{Conclusion}

In this paper, the efficient factorization scheme that we proposed in
a previous study \cite{rosu1} has been applied to more
complicated second order nonlinear differential equations.
Exact particular solutions have been obtained for a number
of important nonlinear differential equations with applications in physics
and biology: the modified Emden equation, the generalized Lienard
equation, the Duffing-van der Pol equation, the convective Fisher
equation, and the generalized Burgers-Huxley equation.



\begin{thebibliography}{99}


\bibitem{rosu1} H.C. Rosu and O. Cornejo-P\'erez,
Phys. Rev. E {\bf 71} (2005) 046607; arXiv:math-ph/0401040

\bibitem{vchandra} V.K. Chandrasekar, M. Senthilvelan, M. Lakshmanan,
Proc. Roy. Soc. Lond. A {\bf 461}  (2005) at press, arXiv:nlin.SI/0408053

\bibitem{chandrasekar} V.K. Chandrasekar, M. Senthilvelan, M. Lakshmanan,
J. Phys. A {\bf 37} (2004) 4527.

\bibitem{schonborn} O. Sch\"{o}nborn, R.C. Desai, D. Stauffer,
J. Phys. A {\bf 27} (1994) L251;
 O. Sch\"{o}nborn, S. Puri, R.C. Desai,
Phys. Rev. E {\bf 49} (1994) 3480.

\bibitem{wang} X.Y. Wang, Z.S. Zhu, Y.K. Lu,
J. Phys. A {\bf 23} (1990) 271.





\end{thebibliography}
\end{document}